\documentclass[twocolumn,showpacs]{revtex4}
\usepackage[xdvi]{graphicx}%
\usepackage{dcolumn}
\usepackage{amsmath}
\makeatletter
\def\btt#1{\texttt{\@backslashchar#1}}%
\DeclareRobustCommand\bblash{\btt{\@backslashchar}}%
\makeatother

\begin{document}
\title{Steady State Thermodynamics of Langevin Systems}
\author{Takahiro Hatano and Shin-ichi Sasa}
\affiliation{Department of Pure and Applied Sciences, 
University of Tokyo, Komaba, Tokyo 153-8902, Japan}
\date{\today}

\begin{abstract}
We study Langevin dynamics describing nonequilibirum steady states.
Employing the phenomenological framework of steady state thermodynamics 
constructed by Oono and Paniconi [Prog. Theor. Phys. Suppl. {\bf130}, 29 (1998)], 
we find that the extended form of the second law which they proposed holds 
for transitions between steady states and that the Shannon entropy difference 
is related to the excess heat produced in an infinitely slow operation.
A generalized version of the Jarzynski work relation 
plays an important role in our theory.
\end{abstract}

\maketitle

The second law of thermodynamics describes the fundamental limitation 
on possible transitions between equilibrium states.
In addition, it leads to the definition of entropy, 
in terms of which the heat capacity and equations of state
can be treated in a unified way. 

In contrast to equilibrium systems, with their elegant theoretical 
framework, the understanding of nonequilibrium steady state systems 
is still primitive. The broad goal with which we are concerned in this 
paper is to establish the connection between the phenomena displayed by 
nonequilibrium steady states and thermodynamic laws. We expect that 
a unified framework that describes both equilibrium and nonequilibrium 
phenomena can be obtained by extending the second law to the state 
space consisting of equilibrium and nonequilibrium steady states.
There have been several attempts to construct such a framework
\cite{jou,keizer,eu,oono}. Among them, a phenomenological framework 
proposed by Oono and Paniconi seems most sophistcated, and their 
framework has been named `steady state thermodynamics' (SST) \cite{oono}.

Oono and Paniconi focused on transitions between steady states 
and distinguished steadily generated heat, 
which is generated even when the system remains in a single state 
in the state space, and the total heat.
They call the former the "house-keeping heat". 
Subtracting the house-keeping heat from the total heat defines 
the excess heat, which reflects the change of the system 
in the state space:
\begin{equation}
\label{Qex}
Q_{\rm ex}\equiv Q_{\rm tot}-Q_{\rm hk}.
\end{equation}
Here $Q_{\rm tot}$ and $Q_{\rm hk}$ denote the total heat and 
the house-keeping heat, respectively.
By convention, we take the sign of heat to be positive 
when it flows from the system to the heat bath.

For equilibrium systems, 
$Q_{\rm ex}$ reduces to the total heat $Q_{\rm tot}$, 
because in this case $Q_{\rm hk}=0$.
Because any proper formulation of SST should reduce to 
equilibrium thermodynamics in the appropriate limit, 
$Q_{\rm ex}$ should correspond to the change of a generalized 
entropy $S$ within the SST.
Here we treat systems in contact with a single heat bath 
whose temperature is denoted by $T$, so that the second law 
of SST reads \cite{oono}
\begin{equation}
\label{secondlaw}
T\Delta S\ge -Q_{\rm ex}.
\end{equation}
The equality here holds for an infinitely slow operation in which 
the system is in a steady state at each time during a transition.
(We call such a process a "slow process".) 
That is, the generalized entropy difference $\Delta S$ 
between two steady states can be measured 
as $-Q_{\rm ex}/T$ resulting from a slow process connecting these two states.
This allows us to define the generalized entropy of nonequilibrium 
steady states, because using it we can determine the generalized entropy 
difference $\Delta S$ between any nonequilibrium steady state and a 
nearby equilibrium state, whose entropy is known.

These are phenomenological considerations and they should ultimately 
be confirmed through experiments.
As a preliminary step toward this confirmation, in this Letter, 
we find support for the validity of the above discussion by studying 
a simple stochastic model.
With the same motivation, Sekimoto and Oono considered 
a simple Langevin system and defined the quantity $Q_{\rm ex}$
\cite{sekimotonote,sekimoto2}. 
However, this nonequilibrium system reduces to an equilibrium system 
through a suitable transformation of variables and hence lacks generality.
Also, one of the present authors has found that 
the minimum work principle holds 
for certain types of transitions between steady states \cite{hatano}, 
with some assumption regarding the steady state measure.
In this Letter, we derive the inequality (\ref{secondlaw}) in 
a more general context and show that the equality holds 
for slow processes.
This result relates the excess heat to the generalized entropy.

We consider the dynamics of a Brownian particle in a circuit 
driven by an external force.
These dynamics are described by the Langevin equation 
\begin{equation}
\label{langevin}
\gamma\dot{x}=-\frac{\partial U(x;\lambda)}{\partial x}
+f+\xi(t),
\end{equation}
where $\xi(t)$ represents Gaussian white noise 
whose intensity is $2\gamma k_{\rm B}T$.
We employ  periodic boundary conditions, and thus 
the particle flows due to the nonconservative force $f$.
This simple nonequilibrium system was investigated by Kurchan 
with regard to the fluctuation theorem \cite{kurchan}.
Transitions between steady states are realized by changing 
the parameters $\lambda$ and $f$.
We assume that if the system is left unperturbed, 
it eventually reaches a steady state which is uniquely 
determined by the parameter values.
Although we consider an explicitly one-dimensional system 
for simplicity, multi-dimensional cases, including 
many-particle systems, are essentially the same.

We write the steady state probability distribution function as 
$\rho_{\rm ss}(x;\alpha)$, where $\alpha$ denotes the set of 
control parameters of the system, $\lambda$ and $f$.
Then we manipulate the system by changing the value of $\alpha$ 
during the interval from $t=0$ to $t=\tau^{\ast}$.
We assume that the system is initially in a steady state, 
and after the completion of the manipulation 
it converges to a new steady state.
Let $\tau$ denote the time at which the system reaches 
the new steady state $(0<\tau^{\ast}<\tau)$.
We descretize $[0,\tau]$ as $[t_0,t_1,\cdots, t_N]$.
We denote the value of $\alpha$ at the $i$th time step by $\alpha_i$. 
This value changes at each time step from time $t_0$ until time 
$t_M=\tau^{\ast}$, while after this time it remains fixed: 
$\alpha_i=\alpha_M$ for $i\ge M$.
We also write $x(t_i)$ as $x_i$.
We consider the limit of an infinitely fine discretization 
by keeping $\tau^{\ast}$ and $\tau$ fixed and taking 
$N\rightarrow\infty$.

Let us introduce a new quantity $\phi(x;\alpha)$ defined by 
\begin{equation}
\label{phi}
\phi(x;\alpha)=-\log\rho_{\rm ss}(x;\alpha), 
\end{equation}
where $\rho_{\rm ss}(x;\alpha)$ is the probability distribution 
function of the steady state corresponding to $\alpha$.
Let $P(x'|x;\alpha)$ be the transition probability 
from $x$ to $x'$ in one time step (whose length is $\Delta t=\tau/N$) 
for a given value of $\alpha$.
Note that by definition 
\begin{equation}
\label{invariant}
\int dxP(x|x';\alpha)\rho_{\rm ss}(x';\alpha)
=\rho_{\rm ss}(x;\alpha).
\end{equation}
Then for a given sequence $(x_0,x_1,\cdots, x_N)$, 
which is collectively denoted by $[x]$, the average of 
a quantity $g([x])$ is written as 
\begin{equation}
\label{pathaverage}
\langle g\rangle\simeq\int dx_N\left[\prod_{i=0}^{N-1}
\int dx_iP(x_{i+1}|x_i;\alpha_{i})
\rho (x_0;\alpha_0)\right] g([x]),
\end{equation}
where the symbol $\simeq$ expresses that this is an approximate equality 
that becomes exact in an appropriate, infinitely fine descretization 
limit of $N\rightarrow\infty$.

Now, in order to derive Eq. (\ref{secondlaw}) 
for the system described by Eq. (\ref{langevin}), 
we utilize a Jarzynski-type equality.
For transitions between isothermal equilibrium states, 
it is known that the following equality holds between the work 
done to the system $W$ and the equilibrium Helmholtz free energy 
difference $\Delta F$ \cite{jarzynski}: 
\begin{equation}
\label{ejarzynski}
\langle e^{-\beta W}\rangle_c =e^{-\beta\Delta F}.
\end{equation}
Here $\beta=1/k_BT$ and $\langle\cdot\rangle_c$ denotes 
the average over all possible histories with respect to 
equilibrium fluctuations.
Note that the minimum work principle $\langle W\rangle_c\ge\Delta F$ 
immediately follows from this relation, due to the Jensen inequality 
$\langle e^x\rangle \ge e^{\langle x\rangle}$.
In a similar way, we now set out to derive Eq. (\ref{secondlaw}) through 
the somewhat generalized version of Eq. (\ref{ejarzynski})
\begin{equation}
\label{gjarzynski2}
\langle\exp[-\beta Q_{\rm ex}-\Delta\phi]\rangle=1, 
\end{equation}
where $\Delta\phi =\phi(x_N;\alpha_N)-\phi(x_0,\alpha_0)$.

We start with the identity 
\begin{equation}
\label{trivial}
\left\langle
\left[\prod_{i=0}^{N-1}\frac{\rho_{\rm ss}(x_{i+1};\alpha_{i+1})}
{\rho_{\rm ss}(x_{i+1},\alpha_i)}\right]
\right\rangle\simeq 1
\end{equation}
This follows from Eqs. (\ref{invariant}) and (\ref{gav}). 
Rewriting Eq. (\ref{trivial}) using $\phi$, we have 
\begin{equation}
\label{trivial2}
\left\langle\exp\left[\sum_{i=0}^{N-1}
\{-\phi(x_{i+1};\alpha_{i+1})+\phi(x_{i+1};\alpha_i)\}
\right]\right\rangle\simeq 1.
\end{equation}
Taking the limit $N\rightarrow\infty$, 
Eq. (\ref{trivial2}) becomes 
\begin{equation}
\label{gjarzynski}
\left\langle\exp\left[-\int_0^{\tau}dt\dot{\alpha}
\frac{\partial\phi (x;\alpha)}{\partial\alpha}\right]
\right\rangle =1.
\end{equation}
It can be easily seen that Eq. (\ref{gjarzynski}) reduces to 
the equilibrium Jarzynski equality (\ref{ejarzynski}) 
when we set $\phi=-\beta(F-U)$.

Now we express the left-hand side of Eq. (\ref{gjarzynski}) 
in terms of heat, so that we can find the correspondence 
with Eq. (\ref{gjarzynski2}). 
First, for Langevin systems, the total heat flowing into 
the heat bath, $Q_{\rm tot}$, is defined by 
\begin{equation}
\label{qtot}
Q_{\rm tot}=\int_0^{\tau}dt[\gamma\dot{x}(t)-\xi(t)]\dot{x}(t).
\end{equation}
Note that the products of ${\dot x}(t)$ and the other quantities 
are of the Stratonovich type.
This interpretation of the heat was proposed and investigated 
by Sekimoto \cite{sekimoto}.
In addition, we note that $\beta Q_{\rm tot}$ 
satisfies the fluctuation theorem 
if the system remains in a steady state \cite{kurchan}.

Next we rewrite Eq. (\ref{langevin}) as  
\begin{equation}
\label{lan1}
\gamma\dot{x}=b(x)-
\beta^{-1}\frac{\partial\phi(x;\alpha)}{\partial x}+\xi(t), 
\end{equation}
where 
\begin{equation}
\label{b}
b(x)=f-\frac{\partial U(x;\alpha)}{\partial x}
+\beta^{-1}\frac{\partial\phi(x;\alpha)}{\partial x}.
\end{equation}
Equation (\ref{lan1}) corresponds to the decomposition 
of the flux ${\dot x}$ into an irreversible part $b(x)$ 
and a reversible part $\partial\phi/\partial x$, 
in the sense of Refs. \cite{graham,eyink}.
Multiplying Eq. (\ref{lan1}) by $\dot{x}(t)dt$ and 
integrating with respect to $t$ from $t=0$ to $t=\tau$, 
we get 
\begin{equation}
\label{lan3}
\beta Q_{\rm tot}=\int_0^{\tau}dt\beta b(x)\dot{x}(t)
-\Delta\phi +\int_0^{\tau}dt\frac{\partial\phi(x;\alpha)}{\partial\alpha}
\dot{\alpha}(t).
\end{equation}
Here we define the house-keeping heat as \cite{footnote1} 
\begin{equation}
\label{Qhk}
Q_{\rm hk}=\int_0^{\tau}dtb(x)\dot{x}(t).
\end{equation}
Using $Q_{\rm ex}=Q_{\rm tot}-Q_{\rm hk}$, 
we can rewrite Eq. (\ref{gjarzynski}) as the generalized 
Jarzynski equality (\ref{gjarzynski2}).

Now we derive the Second Law for SST.
{}From Eq. (\ref{gjarzynski2}) and the Jensen inequality, 
we obtain 
\begin{equation}
\label{inequality}
\beta\langle Q_{\rm ex}\rangle+\langle\Delta\phi\rangle
\ge 0.
\end{equation}
We assume that $\tau$ is sufficiently large, so that 
the correlation between $x_0$ and $x_N$ can be ignored.
Then $\langle\Delta\phi\rangle$ reduces to the difference 
between the averages of $\phi(x;\alpha)$ with respect to 
the initial and the final steady state measures. 
Now, note that with the information-theoretic (Shannon) entropy 
given as 
\begin{equation}
{\cal S}(\alpha)=-\int dx \rho_{\rm ss}(x;\alpha)
\log\rho_{\rm ss}(x;\alpha),
\end{equation}
the quantity $\langle\Delta\phi\rangle$ is equal to $\Delta {\cal S}$.
Equation (\ref{inequality}) then becomes 
\begin{equation}
\label{inequality2}
T\Delta{\cal S}\ge -\langle Q_{\rm ex}\rangle.
\end{equation}
Thus if we identify the Shannon entropy with the generalized entropy 
$S$, we obtain the second law for SST, Eq. (\ref{secondlaw}).

We next consider a slow process in which the distribution 
function of the system can be regarded as $\rho_{\rm ss}(x_i;\alpha_i)$ 
at each time $t_i$. We then have  
\begin{eqnarray}
\left\langle\int d\alpha\frac{\partial\phi(x;\alpha)}
{\partial\alpha}\right\rangle
&=&\int\int d\alpha dx\rho_{\rm ss}(x;\alpha)
\frac{\partial\phi(x;\alpha)}{\partial\alpha}\\
&=&0.
\end{eqnarray}
Recalling that 
\begin{equation}
\int d\alpha\frac{\partial\phi (x;\alpha)}{\partial\alpha} 
=\Delta\phi-Q_{\rm ex},
\end{equation} 
we can prove that the equality holds in Eq. (\ref{inequality2}) 
for a slow process.

Equation (\ref{inequality}), together with the discussion following it, 
constitutes the main result of this Letter.
In the following we discuss five important points 
that are peripherally related with this main result.

First, because we have been able to define a generalized entropy, 
we can also define a generalized Helmholtz free energy $F$ valid for 
nonequilibrium steady states.
\begin{equation}
\label{freeenergy}
F(\alpha)=\int dx\rho_{\rm ss}(x;\alpha)U(x;\lambda)-TS(\alpha).
\end{equation}
{}From the second law for SST, Eq. (\ref{inequality2}), 
if we define the excess work by 
\begin{equation}
W_{\rm ex}=Q_{\rm ex}+\Delta U, 
\end{equation}
the minimum work principle for SST immediately follows: 
\begin{equation}
\label{minimumwork}
\langle W_{\rm ex}\rangle-\Delta F\ge 0.
\end{equation}
The equality here holds for slow processes, 
as in the case of Eq. (\ref{inequality2}).

The next point we wish to discuss regards the function $b(x)$.
The physical meaning of this function as defined by Eq. (\ref{b}) is 
somewhat unclear.
Because this quantity also appears in the definition of the 
house-keeping heat, it would be helpful if we could obtain a more 
intutitive expression for it.
We now derive such an expression.
We consider the local probability current, $j_{\rm ss}$, 
for a given steady state.
\begin{equation}
\label{FP}
\gamma j_{\rm ss}=-\beta^{-1}\frac{\partial\rho_{\rm ss}(x;\alpha)}
{\partial x}+[f-\frac{\partial U(x;\alpha)}{\partial x}]
\rho_{\rm ss}(x;\alpha).
\end{equation}
This value is independent of $x$ for the one-dimensional case.
Using Eqs. (\ref{phi}) and (\ref{b}), Eq. (\ref{FP}) becomes 
\begin{equation}
\label{Qhk2}
b(x)=\frac{\gamma j_{\rm ss}}{\rho_{\rm ss}(x;\alpha)}, 
\end{equation}
which is propotional to the local average velocity.

As the third point of interest, we now discuss the relation 
between the generalized Jarzynski equality 
and the fluctuation theorem 
\cite{evans,lebowitz,maes,kurchan}.
Our argument is the generalization of the Crooks argument 
\cite{crooks}, which focuses on transitions between 
equilibrium states.

We first review the fluctuation theorem, 
following Ref. \cite{maes}. 
Let $\sigma([x])$ be defined according to 
\begin{equation}
\label{deriveft1}
\exp[-\tau\sigma([x])]=\left[\prod_{i=0}^{N-1}
\frac{P(\tilde{x}_{i+1}|\tilde{x}_i;\tilde{\alpha_i})}
{P(x_{i+1}|x_i;\alpha_i)}\right]
\frac{\rho_{\rm ss}(\tilde{x}_0;\tilde{\alpha}_0)}
{\rho_{\rm ss}(x_0;\alpha_0)},
\end{equation}
where $\tilde{x}_i=x_{N-i}$ and 
$\tilde{\alpha_i}=\alpha_{N-i}$.
Note that we can express $\tau\sigma$ as 
\begin{equation}
\label{ft2}
\tau\sigma =\beta Q_{\rm tot}-\Delta\phi 
\end{equation}
by using an explicit form of $P(x_{i+1}|x_i;\alpha_i)$ 
\cite{onsager}.
By a straightforward calculation, 
we find that the probability distribution of $\sigma([x])$, 
which is denoted by $\Pi_{\sigma}(z)$, satisfies
\begin{equation}
\label{ft}
\Pi_{\sigma}(z)=\exp(\tau z)\tilde{\Pi}_{\sigma}(-z),
\end{equation}
where the function $\tilde{\Pi}_{\sigma}$ is the probability distribution 
of $\sigma([x])$ for the system with the parameter set ${\tilde\alpha}$ 
obtained from $\alpha$ under time reversal. 
Equation (\ref{ft}) leads to 
\begin{equation}
\label{ft1}
\langle\exp(-\tau\sigma)\rangle=1.
\end{equation}
We remark that for the case of time-independent $\alpha$, 
Eq. (\ref{ft}) reduces to the relation referred to 
as the fluctuation theorem.

As Crooks demonstrated \cite{crooks}, Eq. (\ref{ft1}) 
yields the Jarzynski equality (\ref{ejarzynski}) 
if we are concerned with transitions between equilibrium states.
However, for transitions between nonequilibrium steady states, 
Eqs. (\ref{ft2}) and (\ref{ft1}) with the Jensen inequality 
do not provide our result Eq. (\ref{gjarzynski2}), 
but rather $T\Delta S\ge -\langle Q_{\rm tot}\rangle$.
Although this inequality does hold, the equality cannot be realized 
since $-\langle Q_{\rm tot}\rangle$ is negative infinite for slow processes.
Thus we cannot define the generalized entropy through Eq. (\ref{ft1}).

In order to clarify the difference between Eq. (\ref{gjarzynski2}) 
and Eq. (\ref{ft1}), we rewrite Eq. (\ref{trivial}) as 
\begin{equation}
\label{dualpath}
\left\langle\frac{\rho_{\rm ss}(x_N;\alpha_N)}{\rho_{\rm ss}(x_1;\alpha_0)}
\left[\prod_{i=1}^{N-1}\frac{P^{\dagger}(x_i|x_{i+1};\alpha_i)}
{P(x_{i+1}|x_i;\alpha_i)}\right]\right\rangle =1,
\end{equation}
where $P^{\dagger}(x_i|x_{i+1};\alpha_i)$ is the dual transition 
probability \cite{prob} defined by 
\begin{equation}
\label{dual}
P^{\dagger}(x_{i}|x_{i+1};\alpha_i)\rho_{\rm ss}(x_{i+1};\alpha_i)
=P(x_{i+1}|x_i;\alpha_i)\rho_{\rm ss}(x_{i};\alpha_i).
\end{equation}
When the detailed balance condition is satisfied, 
Eq. (\ref{dualpath}) together with Eq. (\ref{deriveft1}) 
leads to Eq. (\ref{ft1}), due to the equality 
$P^{\dagger}=P$.
However, the detailed balance condition is violated in  
nonequilibrium steady states, 
and thus the generalized Jarzynski equality 
is not directly related to the fluctuation theorem.

The above discussion leads us to the fourth important point 
we wish to discuss hear, regarding the extent to which 
detailed balance is violated.
For this purpose, we define the quantity $B$ as 
\begin{equation}
\label{B}
\exp[-B(x_{i+1},x_i;\alpha_i)]=
\frac{P(x_i|x_{i+1};\alpha_i)\rho_{\rm ss}(x_{i+1};\alpha_i)}
{P(x_{i+1}|x_{i};\alpha_i)\rho_{\rm ss}(x_{i};\alpha_i)}.
\end{equation}
Then using Eqs. (\ref{gjarzynski2}), (\ref{dualpath}), 
and (\ref{dual}), we obtain an alternative expression of 
the house-keeping heat $Q_{\rm hk}$: 
\begin{equation}
\beta Q_{\rm hk}\simeq\sum_{i=1}^{N-1}B(x_{i+1},x_i;\alpha_i).
\end{equation}

The fifth point we wish to discuss here is the relation between 
our present results and those previously presented by one of 
the authors \cite{hatano}.
Note that if we can decompose $\phi$ as
\begin{equation}
\label{decomposition}
\phi=-\beta[F^{\ast}-\chi(x;f)-U(x;\lambda)], 
\end{equation}
the following equality holds:
\begin{equation}
\label{pastresult}
\left\langle\exp [-\beta\int d\lambda\frac{\partial U(x;\lambda)}
{\partial\lambda}]\right\rangle = \exp[-\beta \Delta F^{\ast}].
\end{equation}
We point out that $F^{\ast}$ here is different from the free energy $F$ 
defined by Eq. (\ref{freeenergy}).
It is seen that Eq. (\ref{pastresult}) is a special case 
of Eq. (\ref{gjarzynski}), since here the only control parameter is 
$\lambda$, and the assumption implicit in the decomposition of 
Eq. (\ref{decomposition}) is nessesary for its derivation.


In conclusion, by defining the excess heat, 
we have derive the second law for SST, Eq. (\ref{inequality2}), 
in the case of a simple stochastic model.
The corresponding thermodynamic function, 
the generalized entropy, is found to be the Shannon entropy.
Also, it is found that for a slow process, the change in this entropy 
is identical to the excess heat divided by the temperature. 

The authors are grateful to Y. Oono for guiding us to the study of SST 
and for enlightening discussions on several related subjects.
They also thank C. Jarzynski and K. Sekimoto for useful discussions 
on the thermodynamics of Langevin systems, and G. C. Paquette for
critical reading of the manuscript. 
This work was supported by JSPS research fellowships, 
and was partially funded by grants from 
the Ministry of Education, Science, Sports and Culture of Japan.


\begin{thebibliography}{99}
\bibitem{footnote1}
This expression of $Q_{\rm hk}$ does not appear in Ref. \cite{oono}.
\end{thebibliography}
\end{document}